# Valence change of praseodymium in $Pr_{0.5}Ca_{0.5}CoO_3$ investigated by x-ray absorption spectroscopy

J. Herrero-Martín[1], J. L. García-Muñoz[1], S. Valencia[2], C. Frontera[1], J. Blasco[3], A.J. Barón-González[1], G. Subías[3], R. Abrudan[4], F. Radu[2], E. Dudzik[2] and R. Feyerherm[2]

[1]*Institut de Ciència de Materials de Barcelona, CSIC, Campus Universitari de Bellaterra, E-08193, Bellaterra, Barcelona, Spain*

[2]*Hemholtz-Zentrum Berlin für Materialen und Energie, Albert-Einstein-Str. 15, 12489 Berlin, Germany*

[3]*Instituto de Ciencia de Materiales de Aragón, CSIC-Universidad de Zaragoza, c/ Pedro Cerbuna 12, 50009 Zaragoza, Spain*

[4]*Institut für Experimentalphysik/Festkörperphysik, Ruhr-Universität Bochum, Germany*

*abstract*

X-ray absorption spectroscopy measurements in $Pr_{0.5}Ca_{0.5}CoO_3$ were performed at the Pr $M_{4,5}$, Pr $L_3$, and Ca $L_{2,3}$ absorption edges as a function of temperature below 300 K. Ca spectra show no changes down to 10 K while a noticeable thermally dependent evolution takes place at the Pr edges across the metal-insulator transition. Spectral changes are analyzed by different methods, including multiple scattering simulations, which provide quantitative details on an electron loss at Pr $4f$ orbitals. We conclude that in the insulating phase a fraction (15($\pm$5)%) of $Pr^{3+}$ undergoes a further oxidation to adopt a hybridized configuration composed of an admixture of atomic-like $4f^1$ states ($Pr^{4+}$) and $f$- symmetry states on the O $2p$ valence band ($Pr^{3+}\underline{L}$ states) indicative of a strong $4f$- $2p$ interaction.

# I. INTRODUCTION

Metal-insulator transitions (MIT) in perovskite cobaltites are a fundamental research topic due to the relevance of the spin state of Co for electron mobility in these strongly correlated oxides [1-8]. In particular, (Pr,$Ln$)$_{1-x}$Ca$_x$CoO$_3$ cobaltites ($Ln$: lanthanide or rare-earth) with perovskite structure are attracting great interest due to their unusual physical behavior [9-16]. As in other heavily doped isostructural cobaltites (such as $Ln_{0.5}$Sr$_{0.5}$CoO$_3$), the metallic state in Pr$_{1-x}$Ca$_x$CoO$_3$ is based on the Co-O e$_g$($\sigma^*$) band, and the electronic configuration of cobalt ions can be described as $t^5_{2g}(\sigma^*)^{1-x}$. Thus, the carriers introduced by doping in Pr$_{0.5}$Ca$_{0.5}$CoO$_3$ can be viewed as low spin (LS) Co$^{4+}$ (S=1/2, $t^5_{2g}$) species moving through the matrix of intermediate spin (IS) Co$^{3+}$ (S=1, $t^5_{2g}e^1_g$) centers.

A sudden spin state transition from IS Co$^{3+}$ to diamagnetic LS Co$^{3+}$ (S=0, $t^6_{2g}$) was proposed as the origin of the MIT in Pr$_{0.5}$Ca$_{0.5}$CoO$_3$ at T$_{MI}$~75 K [9]. In contrast with the ferromagnetic order in compositions that do not show a MIT, the abrupt decrease in the conductivity at T<T$_{MI}$ in Pr$_{0.5}$Ca$_{0.5}$CoO$_3$ avoids the onset of long range order ferromagnetism. Initially this sharp change was solely attributed to the onset of a new electronic configuration of Co (spin state change of Co$^{3+}$ ions). The understanding of the detailed mechanism for electron localization in this system however requires further efforts. Decreasing the $Ln$ size and the Co-O-Co bending angle diminishes the e$_g$($\sigma^*$) bandwidth and the carrier mobility. With $Ln$=Nd or smaller lanthanides in $Ln_{1-x}$Ca$_x$CoO$_3$ (x≤0.5) the conductivity decreases, but there are no signs of abrupt spin state changes [17-20], suggesting only smooth thermally assisted variations of the Co$^{3+}$ spin state. In addition, a first order MIT is only found as a function of temperature or applied pressure in (Pr$_{1-y}Ln_y$)$_{1-x}$Ca$_x$CoO$_3$ compositions having Pr atoms [13,15,17-19].

Recent findings point to an active participation of Pr atoms in the transition [11,13-15]. There is a remarkable volume reduction (~2%) of the unit cell across $T_{MI}$ [10, 11, 21]. Neutron diffraction revealed that this contraction in the insulator phase of $Pr_{0.5}Ca_{0.5}CoO_3$ is the result of an A-O bond shortening (both Pr and Ca occupy the A- site of the perovskite) and it is not due to changes in the size of $CoO_6$ octahedra [11]. In agreement with the theoretical predictions by Knížek *et al* [13], a Schottky anomaly in the specific heat of $(Pr_{0.85}Y_{0.15})_{0.7}Ca_{0.3}CoO_3$ reported by Hetjmanek *et al* indicated the presence of some Kramers $Pr^{4+}$ ions in this compound at low temperature [14]. Finally, x- ray absorption experiments at the Pr $L_3$ and Co $L_{2,3}$ edges of $Pr_{0.5}Ca_{0.5}CoO_3$ allowed García-Muñoz *et al* to confirm the existence of an exceptional mechanism based on a charge migration from Pr to Co at the MIT [15].

Some works have already pointed to a decisive role of Pr-O bonds in this family and a rather unique mechanism for the MIT in heavily Ca-doped Co perovskites that also contain Pr. Nevertheless, a proper understanding of the remarkable electronic and magnetic properties of the reference $Pr_{0.5}Ca_{0.5}CoO_3$ system requires additional investigations. For instance, they should help us to understand the observed photoinduced phase transition in $Pr_{0.5}Ca_{0.5}CoO_3$ under illumination by laser radiation [16], where metallic domains can be stimulated, and which seems to be related to induced spin state changes. It is also important to ascertain the actual occupation of atomic 4*f* states versus other *f*- symmetry extended states not localized on Pr ions, and to clarify the degree of covalency associated to distinct individual Pr-O*i* bonds in the insulating regime.

To obtain more detailed insight on the physical implications associated to the reported (Pr,Ca)-O bond contraction at $T_{MI}$, we report and analyze here the changes occurring at Pr $M_{4,5}$ in the x-ray absorption spectra of $Pr_{0.5}Ca_{0.5}CoO_3$ when undergoing the MIT. Also, we have performed a thorough analysis of the Pr $L_3$ edge results presented in Ref.15, including multiple scattering simulations. We clearly demonstrate that the spectral evolution is related to a further oxidation of $Pr^{3+}$ ions in the insulating phase and make an estimation of the $Pr^{3+}$:$Pr^{4+}$ ratio at 10 K, although the ionic description is too crude in this system where Pr-O covalency effects are crucial. At this temperature, our findings supportthat about a 15% of Pr atoms have lost the trivalent $4f^2$ state, which has been substituted by a more oxidized mixed configuration composed of mainly tetravalent $4f^1$ states but also of a tetravalent Pr with electronic occupation of $f$- symmetry projected extended states on O atoms. In addition, our study of the spectra recorded with hard x-rays rules out intermediate valence models for praseodymium ($Pr^{3+\delta}$) in $Pr_{0.5}Ca_{0.5}CoO_3$.

## II. EXPERIMENTAL

$Pr_{0.5}Ca_{0.5}CoO_3$ (PCCO) samples were prepared in polycrystalline form by solid-state reaction using high purity $Pr_6O_{11}$, $Co_3O_4$ and $CaCO_3$ precursors, following the method described in Ref.11. The final annealing temperature was 1160 °C under $O_2$ atmosphere. To reach the optimal oxygen content, the samples were also treated under high oxygen pressure (at 900ºC with $P_{O2}$=200 bar during 14 hours, and at 475ºC with $P_{O2}$=150 bar during 6 hours). Samples were single phased, and fully stoichiometric according to Rietveld analysis of neutron diffraction data. Resistivity and magnetization curves displayed an abrupt transition signaling that electron localization occurs at $T_{MI}$~ 75 K [11].

X-ray absorption spectra (XAS) measurements were made at the synchrotron radiation source of the Helmholtz Zentrum in Berlin [22]. Spectra were collected in beamlines MAGS (Pr $L_3$ edge) and PM3 (Pr $M_{4,5}$ and Ca $L_{2,3}$ edges) by means of bulk-sensitive Fluorescence Yield (FY) and surface sensitive Total Electron Yield (TEY). MAGS beamline provides a focused beam in the energy range 4–30 keV, monochromatized by a Si (111) double crystal monochromator, with a photon flux of the order of $10^{12}$ photons/s at 10 keV. The energy resolution is of the order of 2 eV at 6 keV. PM3 dipole beamline delivers radiation within the 20-1900 eV energy range. Circular incident light polarization was employed. The measured maximum flux is about $10^{10}$ photons/s at 900 eV, and the horizontal and vertical divergences of the beam are 1.5 and 1.0 mrad, respectively. The energy resolution is better than 0.3 eV at 900 eV. The $Pr_{0.5}Ca_{0.5}CoO_3$ polycrystalline sample was placed either in a displex-type (MAGS) or a He flow (PM3) cryostat that allowed temperature dependent measurements. Spectra were collected on heating between 10 K and 300 K. Fluorescence photons were detected either by a scintillation counter (MAGS) or a photodiode (PM3) placed at an angle of ca. 90° with respect to the incoming beam and vertically oriented with respect to the sample. Data normalization was performed following standardized procedures in all cases. Background removal was performed by subtraction of a straight line. At the Pr $L_3$ edge, absorption was normalized to an arbitrary value in the extended energy region, about 50 eV beyond the peak (*i.e.* the white line) position. In Pr $M_{4,5}$ spectra, we normalized to unity at the maximum of the $M_5$ edge

## III. RESULTS AND DISCUSSION

### *III.1 Soft x-rays spectra. Pr $M_{4,5}$ and Ca $L_{2,3}$ edges across the MIT*

Top panel of Fig.1 shows the experimental TEY detected absorption spectra (Pr $M_{4,5}$ edge) of PCCO at different temperatures between 10 K and 300 K. A clear evolution is observed. When crossing the MIT (from high to low temperature), the spectra appear slightly displaced towards higher energy values by up to about 0.15 eV (Fig.1, top, middle: circles/left scale). In addition, the spectral shape evolves (Fig.1, top). *B* and *F* features appearing at $T<T_{MI}$ are characteristic of compounds containing $Pr^{4+}$. As explained within the single impurity Anderson Model, they originate due to the presence of a ligand-hole ($\underline{L}$) in the 2*p* states of nearest neighbouring oxygen atoms [23-25]. Indeed, an atomic-like $4f^1$ state seems to be energetically unfavourable.

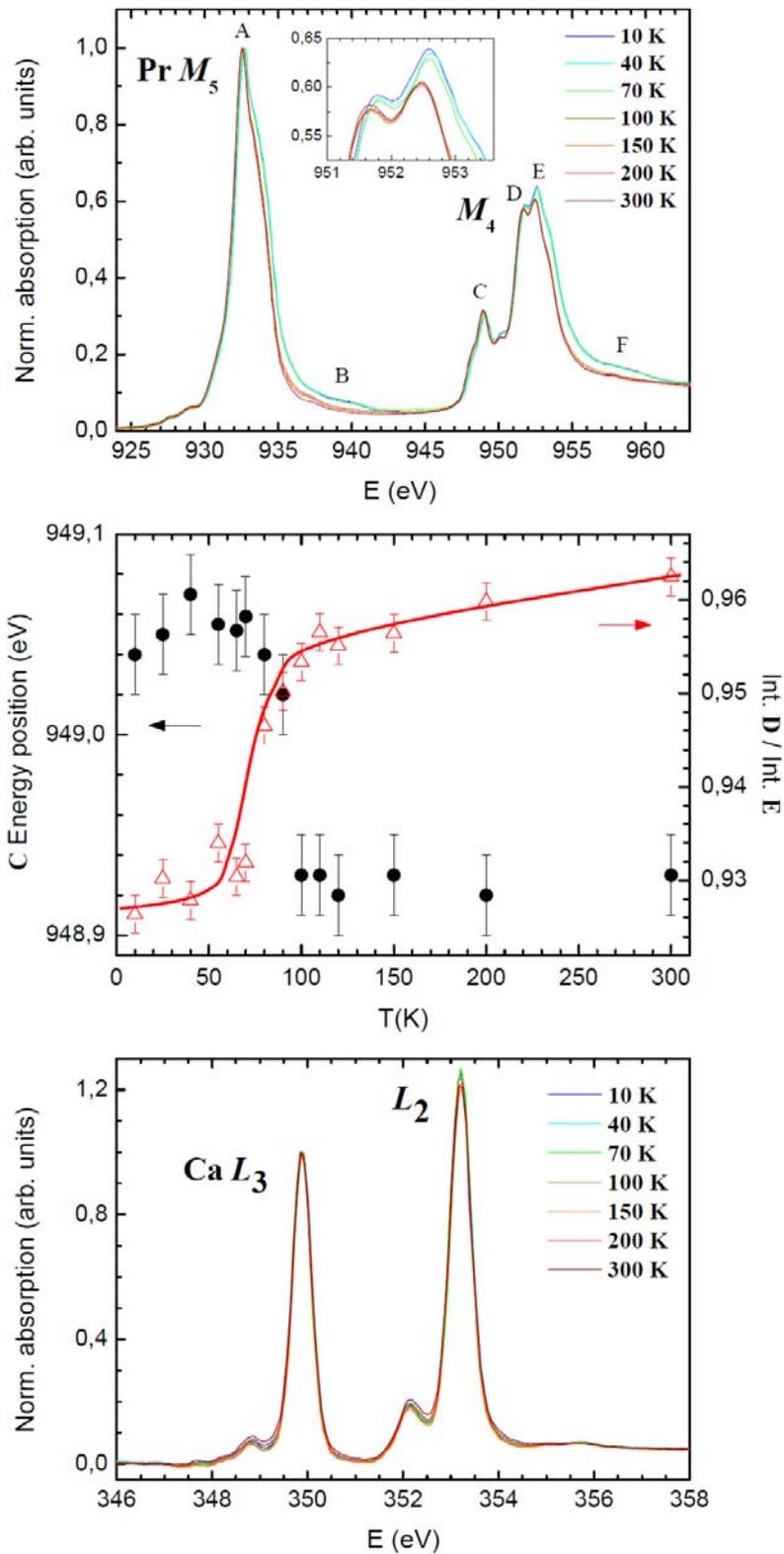

**Figure 1**. (color online) Experimental soft x-ray absorption in $Pr_{0.5}Ca_{0.5}CoO_3$. *Top*: spectra at the Pr $M_{4,5}$ edges as a function of T. A clear shift in the energy position and intensities of *B*, *E* and *F* features occurs at the MIT as shown in the inset; *middle*: fingerprints of the Pr valence change at the MIT such as the energy position of *C* (circles), and *D/E* relation of intensities (triangles), line is a guide to the eye; *bottom*: spectra at the Ca $L_{2,3}$ edges as a function of T.

Covalency effects are large, and the spectroscopic features of nominally $Pr^{4+}$ reference compounds like $PrO_2$ are only understood by considering a significant charge-transfer from oxygen atoms. It has been reported that the combination of pure $Pr^{4+}$ ($4f^1$) and a nominally less oxidized Pr cation associated to a ligand-hole ($4f^2\underline{L}$) leads to a qualitatively correct reproduction of experimental spectra. Actually, the latter component is often nearly as important as the atomic-like one [25-27]. The partial conversion of $Pr^{3+}$ into $Pr^{4+}$ at the MIT produces an evolution in the relative intensities of *D* and *E* peaks too, as seen in the middle panel of Fig.1 (triangles/right scale). These changes seem to make a sense after inspection of the absorption spectra of $Pr_2O_3$ ($Pr^{3+}$) and $BaPrO_3$ ($Pr^{4+}$) reference compounds reproduced from Hu *et al* [28] in Fig.2, which show remarkable differences. In the latter, the maximum at the Pr $M_5$ is found at 934 eV, about 1.5 eV beyond the Pr $M_5$ peak in PCCO at 300 K, and it shows a very large broad maximum at the Pr $M_4$ edge, also shifted to higher energies with respect to the multiple-featured edge in PCCO. The substitution of $Pr^{3+}$ by $Pr^{4+}$ therefore produces a decay in the *D/E* amplitude ratio in the insulating phase.

The bottom panel of Fig.1 shows the absorption spectra of PCCO at the Ca $L_{2,3}$ edges as a function of temperature. It is worth reminding that Pr and Ca atoms randomly occupy the A- position of the perovskite. We expect changes in Pr-O bonds not to be accompanied by a variation in the electronic state of Ca atoms. Indeed, we do not observe any significant change in the spectral shape between 10 and 300 K. Thus, we can safely ascribe the changes reported in the structural environment around the A- site of the PCCO perovskite to Pr atoms only [11].

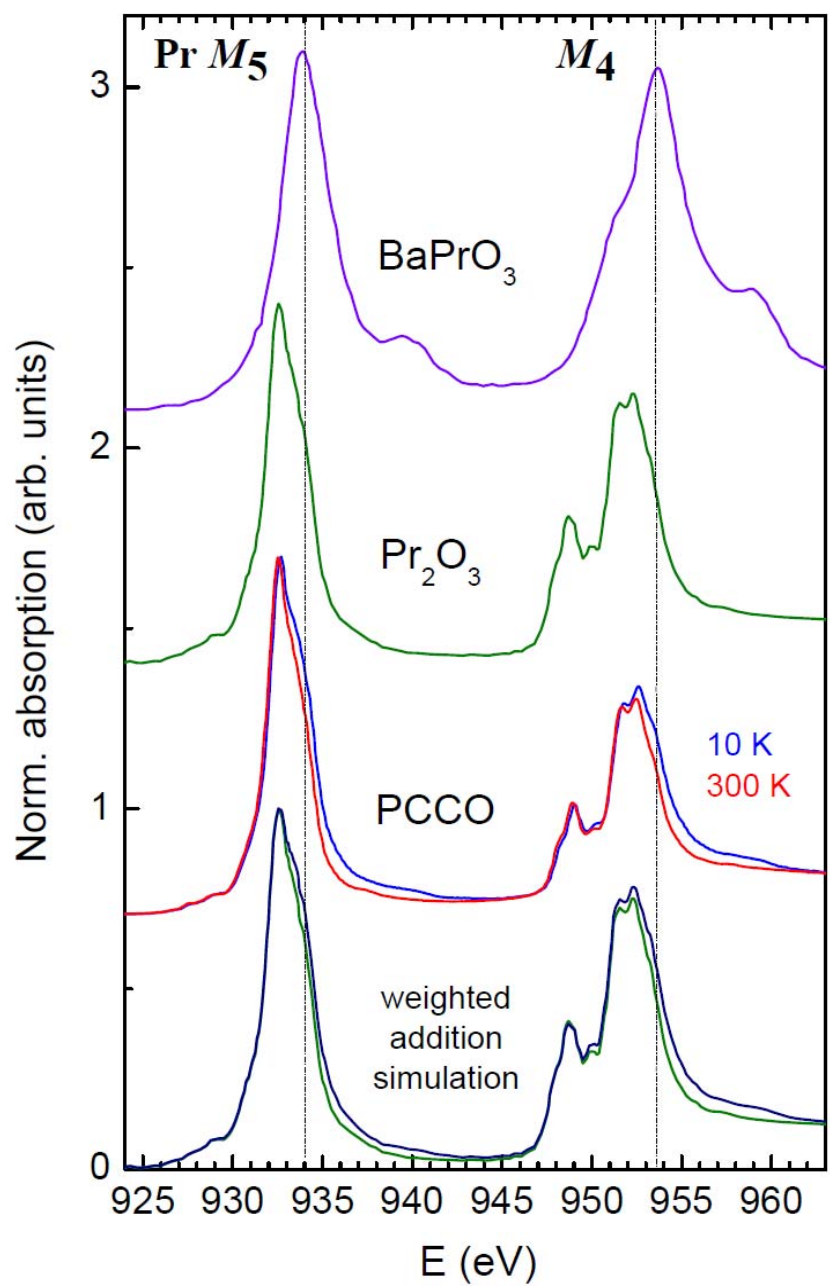

**Figure 2**. (color online) From top to bottom: experimental Pr $M_{4,5}$ absorption spectra of Pr tetravalent BaPrO$_3$ (violet) and trivalent Pr$_2$O$_3$ (green) [both from Ref.28], Pr$_{0.5}$Ca$_{0.5}$CoO$_3$ measured at 10 K (blue) and 300 K (red), and a comparison of Pr$_2$O$_3$ to the 85:15 weighted

addition (dark blue) of $Pr_2O_3$ and $BaPrO_3$ spectra, *i.e.* ideally the spectrum of a compound containing in average $Pr^{+3.15}$. Some spectra have been vertically shifted for clarity reasons.

In the following, we attempt to estimate the actual charge donation by $Pr^{3+}$ cations across the MIT. We start by assuming that no $Pr^{4+}$ is present in our sample at room temperature. In Fig.2, we can thus align the spectrum of PCCO at 300 K with that of $Pr_2O_3$ from Ref.28 in the energy axis (at the Pr $M_5$ peak). Little misalignment detected in the energy position at the Pr $M_4$ edge is most likely due to slight differences in the spin-orbit coupling strength in both compounds. Indeed, PCCO at 300 K and $Pr_2O_3$ spectra are remarkably alike. It is worth noticing that, contrarily to hard x-rays absorption where the electronic band structure and the local geometry are the fundamental parameters modeling the spectral shape, the core-hole related effects (charge transfer, spin-orbit, and exchange and multiplet interactions) dominate in the soft x-rays regime and push crystal structure effects into the background [29].

The weighted addition of $Pr_2O_3$ and $BaPrO_3$ spectra can thus be used to simulate the effect of an oxidation state of Pr ions between +3 and +4 in PCCO at $T<T_{MI}$. In the bottom of Fig.2 we plot the result for the best weighted addition of these two references (85:15). The sum spectrum shows an increase of amplitude in *B* and *F* features comparable to that observed between the experimental spectra in PCCO at 300 K and 10 K. We also see a decrease in the *D*/*E* amplitude ratio, although less noticeable than in PCCO. In general, the sum spectrum becomes a good simulation of the 10 K spectrum of PCCO. Minor discrepancies may be related to the approximation of BaPrO3 as a $Pr^{4+}$ reference. It is known that Pr 4*f*- O 2*p* orbital hybridization effects can strongly modulate Pr $M_{4,5}$ absorption spectra and the amplitude of *B* and *F* features is one of its fingerprints [27,29]. Nevertheless, they have been found to be similar in compounds with a well differentiated structure at room temperature, such as $PrO_2$ (*Fm-* $3m$) and BaPrO$_3$ (*Imma*) [28]. Therefore we can expect a comparable spectral contribution of covalency effects in PCCO and BaPrO$_3$, both with a perovskite structure. Consequently, taking into account the uncertainties inherent to the method employed to analyze Pr $M_{4,5}$ absorption data, we can conservatively estimate about a 15($\pm$5) % of Pr$^{3+}$ ions undergoing a further oxidation process by crossing the MIT. This value is not far but somewhat lower than the estimation in Ref. 15 for the same compound, and below the calculated amount of oxidized Pr$^{4+}$ in (Pr$_{0.85}$Y$_{0.15}$)$_{0.7}$Ca$_{0.3}$CoO$_3$ from specific heat measurements [14].

### *III.2 Hard x-rays spectra. Pr L$_3$ edge results and analysis.*

Despite the presented absorption study at the Pr $M_{4,5}$ ($3d \rightarrow 4f$) edge seems to be a more direct way to look into the actual oxidation state of Pr atoms, the eventual presence of Pr$^{4+}$ is also reflected at the Pr $L_3$ ($2p_{3/2} \rightarrow 5d$) edge [15,30-34].

Top of Fig. 3 shows the experimental spectra at 10 K and 300 K. The Pr$^{4+}$ fingerprint feature is labelled as *S*. In Ref.15 we attributed the enhancement of *S* in PCCO absorption spectra at T<T$_{MI}$ to the appearance of Pr$^{4+}$ ions. However, to further confirm this interpretation it is necessary to analyze in detail the origin of this feature since multiple scattering (MS) of the photoelectron with the locally surrounding atoms around the central absorbing Pr may be significant at this energy value. Moreover, the amplitude of this likely contribution might vary with temperature. Regarding these concerns, on the one hand we must take into account that *S* is observed with different intensities in the metallic regime of PCCO (at high temperature, where praseodymium is present as Pr$^{3+}$) and in other Pr$^{3+}$ reference compounds [15,30,35]. Besides, the atomic displacements observed at T$_{MI}$ induce geometrical variations that could influence the MS.

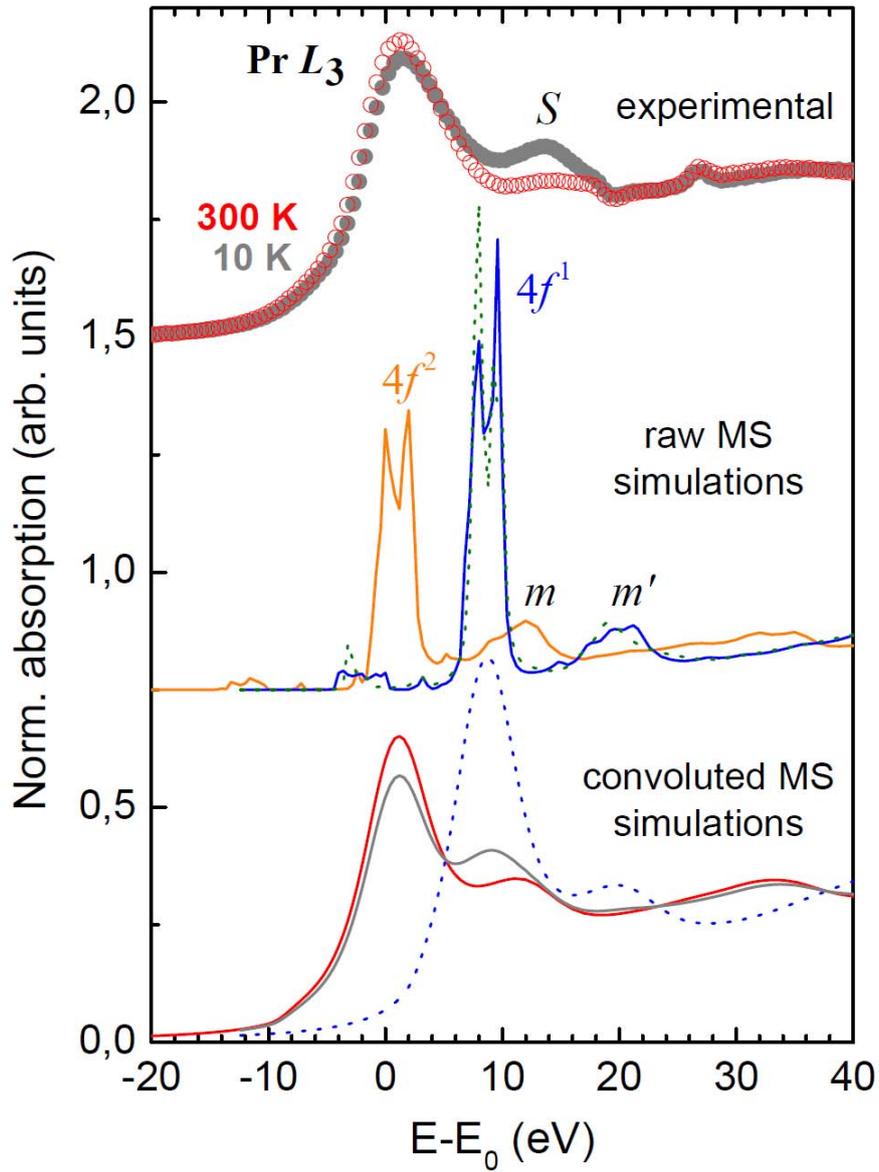

**Figure 3**. (color online) *Top*: experimental absorption spectra at the Pr $L_3$ edge of PCCO at 10 K (grey filled circles) and 300 K (red open circles); *middle*: FDMNES calculations at the Pr $L_3$ edge before convolution by the core-hole energy broadening by using a cluster with radius 6.1 Å around the central Pr atom and a $4f^2$ (orange solid line) and $4f^1$ (blue solid line) electronic configurations, and for a 4.0 Å cluster (27 atoms) with a $4f^1$ configuration (green dotted line); *bottom*: convoluted FDMNES calculated spectra for the larger cluster and a $4f^2$ (red) and $4f^1$ (blue) electronic filling. The grey line is the result of a 85:15 weighted addition of the latter, respectively. Some spectra have been vertically shifted for clarity.

To get a deeper insight on the Pr $L_3$ edge changes, we have performed x-ray absorption simulations based on the MS formalism by means of the FDMNES code [36]. Calculations are relativistic and make use of the muffin-tin approximation for the calculus of the Hedin-Lundqvist potentials. The energy axis is relative to $E_0$, the calculated absorption edge energy for the $4f^2$ electronic configuration. The experimental spectrum has been shifted to align the Pr $L_3$ white line with that of calculations to facilitate comparisons. In the middle part of Fig. 3 we show the result of simulations for a cluster of radius 6.1 Å around Pr atoms prior to convolution by the natural core-hole and experimental energy widths. Simulations have been performed for a PrCoO$_3$ cluster where we have used the crystallographic structure of PCCO at T>T$_{MI}$ for a Pr $4f^2$ electronic configuration, and that of the low temperature insulator phase for an excited Pr $4f^1$ configuration [11]. The inclusion of Ca atoms in the cluster at the same crystallographic positions of Pr was discarded due to computational limitations. Nevertheless, we checked for smaller clusters that the Ca presence barely influences the position and amplitude of the main spectral features at the Pr $L_3$ edge. The MS simulation for the excited state ($4f^1$, Pr$^{4+}$) shows a larger, shifted white line by about 8 eV to a higher energy with respect to the Pr $4f^2$ state simulation. The feature arising from MS is present and shows a similar magnitude (*m*, *m'* in the central panel of Fig.3) in both simulations. Moreover, we have obtained that the simulation of MS for the $4f^1$ state shows the same amplitude as for a smaller cluster containing only 27 atoms. Hence, we must conclude that the enhancement of *S* below the MIT in the experimental spectra cannot be attributed to changes in the MS signal associated to the reported structural variation at T~75 K, but it has to be the consequence of the occupation of excited $4f^1$ states, *i.e.* to the appearance of Pr$^{4+}$.

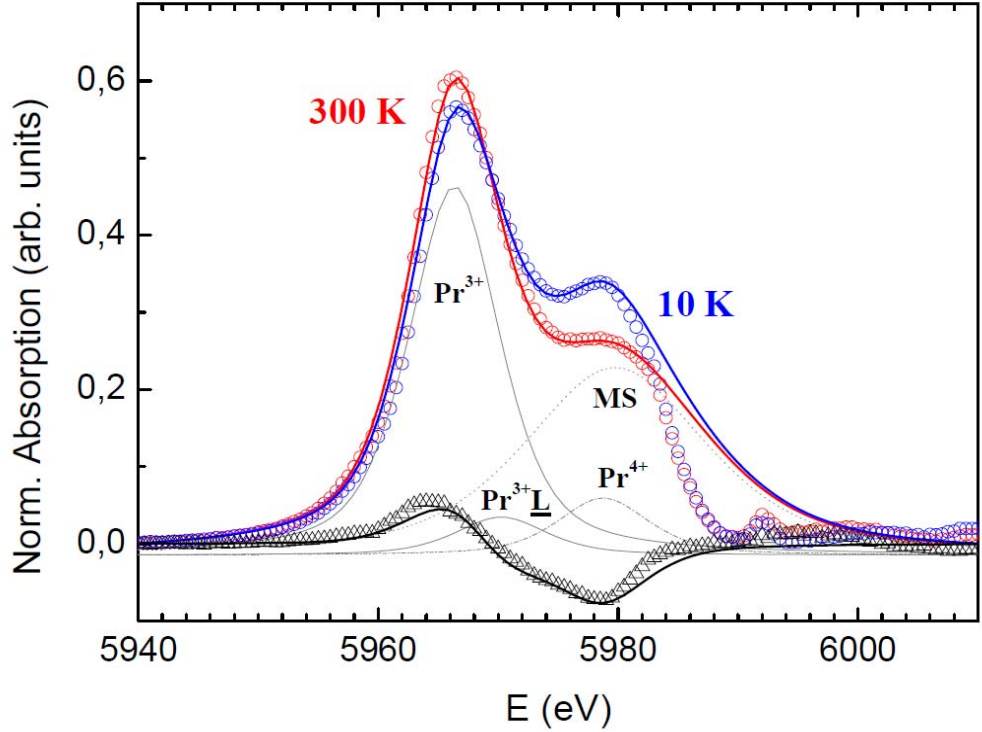

**Figure 4**. (color online) Experimental (circles) and Pseudo-Voigt-composed best fitting curves (solid lines) of Pr $L_3$ absorption spectra. Individual Pseudo-Voigt curves contributing to the spectrum at 10 K are also shown (grey lines). Black triangles and black line are the result of the subtraction of 300 K- 10 K experimental spectra and fitting curves, respectively.

Actually, the so-called *S* feature is formed by two components: the MS and the $Pr^{4+}$ contributions, close in the energy scale. In the previous section we estimated the presence of ~15% of $Pr^{4+}$ at 10 K in PCCO from Pr $M_{4,5}$ experimental absorption data. Next we will check the consistency of that result against the Pr $L_3$ edge data, considering it as an input for the FDMNES simulations. So, in the bottom part of Fig.3 we show the calculated Pr $L_3$ absorption spectra after energy convolution. The grey curve depicts the result of the weighted addition of $4f^2$ and $4f^1$ states calculations according to the ratio 0.85:0.15. The comparison to experimental spectra in the top of Fig.3 shows a good qualitative agreement. The low temperature spectrum with average $Pr^{3.15+}$ content reproduces (*i*) the enhancement of *S* in a nearly quantitative manner ($S_{10K}/S_{300K}$=1.21 and 1.17 for the experimental and simulated cases, respectively), (*ii*) the slight displacement towards lower energy, and (*iii*) the decrease in the amplitude of the main white line and the small shift of the main edge towards a higher energy by about 0.4 eV. There is a small discrepancy in the overall energy position of features in the calculated spectra with respect to experimental ones but it does not affect the basic grounds of our analysis. It is important to notice that the transfer of spectral amplitude from the white line to $S$ at T<$T_{MI}$ would be not easily explainable under a picture where the low temperature phase was described in terms of an intermediate $Pr^{3+\delta+}$ valence state.

More important is to note that the white line decrease at 10 K is overestimated in the calculations performed (Fig.3). We consider that the explanation likely resides in the ligand-hole contribution. This can be due to covalency effects and *f*- electrons delocalization due to strong interaction between Pr 4*f* and O 2*p* states. In many formally tetravalent rare-earth ($Pr^{4+}$, $Ce^{4+}$, $Tb^{4+}$) based compounds covalency promotes the presence of ligand-holes in the neighboring cations and therefore the coexistence of $4f^n$ and $4f^{n+1}\underline{L}$ states [31,32]. Fig.3 suggest that, beyond the atomic orbital description, one should also consider Pr 4*f* states that present large hybridization with O 2*p* states and produce occupation of $4f^2\underline{L}$ states. We have analyzed Pr $L_3$ spectra performing a fit of the absorption spectra by means of Pseudo-Voigt (PV) curves after removal of the spectroscopic contribution of high energy continuum states by subtraction of an arctangent curve [26,32]. A key point in this analysis is to reduce at maximum the number of free parameters. We have employed one PV curve per Pr final state electronic configuration. Thus, at 300 K, assuming Pr valence is purely 3, we used one PV curve to account for the white line centered at 5966.5 eV and a second one, broader to simulate the spectral weight of the secondary multiple scattering feature centered at 5979.7 eV. By fixing the amplitude of the latter PV, we diminished the spectral weight of the first curve to simulate the loss of $Pr^{3+}$ in the 10 K spectrum. This is reintroduced into the spectrum as a gain in $Pr^{4+}$ and $Pr^{3+}\underline{L}$, which are simulated by the addition of two new PV curves (with the same full width at half maximum as that corresponding to $Pr^{3+}$) to the spectrum centered at 5978.7 eV and 5970.2 eV, respectively [26]. The best fitting curves to experimental spectra at 300 K and 10 K are shown in Fig. 4, where 0.15e- per Pr atom have been transferred to produce a mixed state with 0.09(2)e- of pure $Pr^{4+}$ and 0.06(2)e- of $Pr^{3+}\underline{L}$. Namely, a 15($\pm$5) % of Pr atoms are further oxidized at low temperature into a final state $|\psi_{oxid}>=\alpha|4f^1> + \beta|4f^2\underline{L}>$ with $|\alpha|^2 \sim 0.6$ and $|\beta|^2 \sim 0.4$. The large disagreement between experimental and simulated spectra at energies beyond 5982 eV is due to the arctangent truncation function employed. In the bottom part of Fig.4 we can see the experimental and calculated difference spectra between high and low temperature, where the ligand-hole spectral component accounts for the little kink present at about 5970 eV.

## IV. CONCLUSIONS

We have presented a comprehensive spectroscopic study of the thermal evolution of Pr electronic properties in the structurally simple perovskite $Pr_{0.5}Ca_{0.5}CoO_3$. X-ray absorption spectra at both the Pr $M_{4,5}$ and $L_3$ edges undergo clear changes across the metal-insulator transition. These can be correlated to the anomalous shortening of the mean (Pr,Ca)-O distance in the insulating state as observed by neutron diffraction [11]. We have also shown that Ca $L_{2,3}$ spectra do not display perceptible changes across the transition and hence we conclude that only Pr atoms account for this bondlength reduction. The magnitude of the Pr-O1 (-2.4%) and Pr-O2 (-2.1%) distances contraction [15] suggests a large hybridization between Pr 4*f* states wavefunctions and oxygen orbitals and/or electron migration involving changes in the valence of Pr ions.

Analysis of experimental spectra, performed following several methods, demonstrates that the spectral evolution is associated to a partial oxidation of $Pr^{3+}$ ions into $Pr^{4+}$. The low temperature spectrum of PCCO at the Pr $M_{4,5}$ edges could be well reproduced by a semi-empirical simulation making use of nominally $Pr^{3+}$ and $Pr^{4+}$ reference compounds in a 85:15 ratio. Pr $L_3$ edge experimental spectra at both sides of the transition were analyzed taking into account the multiple scattering resonance and the influence of the structural modifications at the transition. Within these considerations, the Pr $L_3$ spectrum at 10 K phase supports the coexistence of roughly a 85:15 admixture of (*i*) localized $4f^2$ states ($Pr^{3+}$), and (*ii*) further oxidized $Pr^{4+}$ atoms. As usually found in formal $Pr^{4+}$ compounds, a high degree of covalency is expected in the second type of Pr atoms, which can be modelled through a ligand-hole in oxygen atoms. Our simulations indicate that nominally $Pr^{4+}$ atoms can be described by an admixture configuration of atomic-like $4f^1$ states (~60%) and $4f^1$ states with a large degree of covalent mixing with O 2*p* orbitals that induce the occupation of *f*-symmetry states localized on the valence band of O atoms ($Pr^{3+}\underline{L}$ states, ~40%), respectively. Indeed, this non-ionic contribution needs to be considered in order to correctly account for some features in the experimental spectra.

Our analysis supports the charge transfer interpretation of the changes detected at the metal-insulator transition, and are consistent with density functional theory (DFT)-type calculations that indicate a transfer of holes from cobalt to praseodymium sites [13], which would promote a Co spin state transition to the LS $t_{2g}^6$ configuration. As a result of the charge transfer, the insulating state seems due to the substitution of a broad itinerant Co-O band by more narrow and localized Pr-O states at the Fermi level.

**Acknowledgements**

We thank financial support from MICINN (Spanish government) under projects MAT2006-11080-C02-02 and No. MAT2009-09308, and NANOSELECT under Project No. CSD2007-00041. We acknowledge ILL (and the CRG-D1B), and HZB for the provision of beamtime. The ALICE diffractometer is funded through the BMBF under Contract No. 05KS7PC1. The research leading to these results has received funding from the European Community´s Seventh Framework Programme (FP7/2007-2013) under grant agreement n° 226716. JHM thanks CSIC for JAEdoc contract and Y. Joly for valuable comments.